# Dephasing of Strong-Field-Driven Floquet States Revealed by Time- and Spectrum-Resolved Quantum-Path Interferometry

Yaxin Liu[1,§], Bingbing Zhu[1,§], Shicheng Jiang[2,§,‡], Shenyang Huang[1,6], Mingyan Luo[1], Sheng Zhang[1], Hugen Yan[1], Yuanbo Zhang[1,4,5], Ruifeng Lu[3,†], Zhensheng Tao[1,*]

[1]*State Key Laboratory of Surface Physics and Key Laboratory of Micro and Nano Photonic Structures (MOE), Department of Physics, Fudan University, Shanghai 200433, P R China*
[2]*State Key Laboratory of Precision Spectroscopy, East China Normal University, Shanghai 200062, P R China*
[3]*Institute of Ultrafast Optical Physics, MIIT Key Laboratory of Semiconductor Microstructure and Quantum Sensing, Department of Applied Physics, Nanjing University of Science and Technology, Nanjing 210094, P R China*
[4]*Shanghai Qi Zhi Institute, Shanghai 200232, P R China*
[5]*New Cornerstone Science Laboratory, Shenzhen 518054, P R China*
[6]*Institute of Optoelectronics, Fudan University, Shanghai 200433, P R China*
[§]These authors contributed equally to this work.

**ABSTRACT**: Floquet engineering, while a powerful tool for ultrafast quantum-state manipulation, faces challenges under strong-field conditions, as recent high harmonic generation studies unveil exceptionally short dephasing times. In this study, using time- and spectrum-resolved quantum-path interferometry, we investigate the dephasing mechanisms of terahertz-driven excitons. Our results reveal a dramatic increase in exciton dephasing rate beyond a threshold field strength, indicating exciton dissociation as the primary dephasing mechanism. Importantly, we demonstrate long dephasing times of strong-field-dressed excitons, supporting coherent strong-field manipulation of quantum materials.

[‡] scjiang@lps.ecnu.edu.cn

[†] rflu@njust.edu.cn

[*] ZhenshengTao@fudan.edu.cn

*Introduction.* – The ability to manipulate material properties at ultrafast speed is essential for advancing photonics technology and enables applications in optical signal processing, on-chip optical sources, and computing. Floquet engineering, which aims to tailor material properties by periodic driving fields [1–4], offers a nonthermal and widely applicable strategy to dynamically control materials. Periodically driven materials exhibit fascinating tailored properties, including modification of topological states [2,5–9], emergence of dynamical localization [10,11], strong-field-driven exciton dynamics [12–14], and modulation of optical properties [15–17]. Recent demonstrations include direct band manipulation of semiconductors [18].

Success of Floquet engineering in solids relies on maintaining coherence of dressed quantum states for a sufficiently long timescale, facilitating further computation or device operations. However, the interaction with the environment in solids inevitably introduces decoherence. Therefore, investigating the possible persistence of Floquet phenomena in the presence of dissipation [19–22] is crucial for future applications. Furthermore, proper dissipation management could help stabilize novel Floquet phases like dissipative time crystals [23–25] and control exotic topological phases [20].

Previous investigations on laser-driven quantum materials have considered the roles of phonon and many-body environments in dephasing, yielding typical dephasing times ($T_2$) of 0.1-1 ps and 10-100 fs [21,22,26,27], respectively. While strong dressing fields are advantageous for inducing giant modulation on material properties [17], recent studies on high-order harmonic generation (HHG) and side-band generation (HSG) from solids have cast doubt on the dephasing mechanism under strong driving fields. State-of-the-art quantum mechanical models have required surprisingly short $T_2$ of 0.1 - 4 fs to reproduce the HHG and HSG spectra [28–33], much shorter than a single cycle of driving fields. Although the short dephasing time was attributed to enhanced electron scattering at high momentum ($k$) [29,33], it remains a subject of debate [34,35].

It is important to note that the dephasing rates from previous experiments were often determined by comparing the decay of harmonic spectra with simulations assuming constant or $k$-dependent dephasing rates [28,29,32]. However, this approach may be inadequate, as contributions from electrons and holes in high-energy bands can obscure the actual dephasing dynamics of laser-dressed, low-energy charge carriers that are responsible for laser-driven coherent currents. It is worth mentioning that more rigorous calculations have shown a field-strength-dependent dephasing time, originating from dynamical Coulomb scattering [36]. To gain insight into the dephasing mechanism of strong-field-driven quantum states, it is crucial to distinguish the signal of low-energy electronic excitations, such as excitons, when measuring the coherence of laser-induced dynamics.

In this work, we investigate exciton dephasing in bulk $MoS_2$ driven by strong terahertz fields using quantum-path interferometry in the framework of Floquet theory. The contribution from the excitonic states is clearly distinguished by the spectrum-resolving capability of our method. By driving the material with a photon energy far below the bandgap, we observe hybridization of exciton 1s and 2p states. The resonant transitions to these dressed states result in amplitude and phase modulations in the spectro-temporal interferograms. We find that the dephasing rate of dressed excitons is field-strength dependent, with its zero-field value consistent with the relaxation pathway of exciton-phonon interaction. Remarkably, under the high-field conditions, we observe a pronounced increase in the dephasing rate, when the field strength exceeds a threshold value of approximately 2.1 MV/cm. This allows us to identify that exciton dissociation plays a dominant role. Our results also show that the dephasing time of a strong-field-driven exciton can exceed 100 fs, even in the non-perturbative region. Our results support the feasibility of strong-field Floquet engineering for low-energy electronic states in quantum materials.

*Dressed excitonic states in bulk MoS$_2$.* – As shown in Fig. 1(a), the experiment involves exciting a 50-nm thick MoS$_2$ with a strong multicycle terahertz (THz) pulse ($F_{THz}$, red waveform) featuring a center frequency of 21 THz and a photon energy of $\hbar\Omega$ =86.8 meV. The peak electric field inside the material can reach 5.6 MV/cm, with its polarization along the Γ-K direction of the Brillouin zone. The THz-pulse excitation coherently forms dressed excitonic states [inset of Fig. 1(a)], where $|0\rangle$ is the vacuum state, and $|1s\rangle$ and $|2p\rangle$ represent the ground and excited states of the Wannier-Mott excitons, respectively. A near-infrared (NIR) probe pulse ($F_{pr}$, blue pulse) with varying frequencies and bandwidths induces phase-locked high-order sideband emission (HSE) [12,33,37], denoted as $|\omega + n\Omega\rangle$, where $\omega$ is the probe-pulse frequency and $n$ is the sideband order. The details of the experimental setup and sample preparation are summarized Supplementary Materials (SM) Section S1.

In Fig. 1(b), we show the absorbance, $\alpha(\omega, F_{THz})$, of MoS$_2$ with and without the THz-pump excitation, probing the dressed quantum system with a broadband NIR pulse (see SM Section S3). Under the field-free condition, the resonant feature at 448 THz (~1.85 eV) can be attributed to the transition energy of $|1s\rangle$ of the type-A exciton ($E_{XA}$) [38,39]. Another absorption feature at a higher energy of ~496 THz (~2.05 eV) is attributed to the type-B exciton ($E_{XB}$) [38,39]. For this work, we focus on the type-A exciton due to the NIR-pulse bandwidth. The THz field induces strong absorbance change, including the reduction and broadening of the absorption peak and a blueshift in energy.

Because the dressing THz field oscillates at a frequency significantly lower than the bandgap, it strongly couples the $|1s\rangle$ and $|2p\rangle$ excitonic states. According to the Floquet theorem [3,8], the dressed excitonic state is a superposition of equidistant energy state manifolds, containing two states, $|1s, l\rangle$ and $|2p, l-1\rangle$, in each Floquet Brillouin zone (FBZ), where $l$ is an integer ($l \in Z$) denoting the number of dressing photons, and the FBZ width is $\hbar\Omega$ [inset of Fig. 1(a)]. The hybridization of $|1s, l\rangle$ and $|2p, l-1\rangle$ in a single FBZ can be described by [17]

$$|1s, l\rangle = e^{-i[\varepsilon_{1s}+l\hbar\Omega+\Delta E]t/\hbar}\left(-\sin\beta\, e^{-i\varphi}|2p\rangle + \cos\beta\,|1s\rangle\right), \qquad (1)$$

$$|2p, l-1\rangle = e^{-i[\varepsilon_{2p}+(l-1)\hbar\Omega-\Delta E]t/\hbar}\left(\cos\beta\,|2p\rangle + \sin\beta\, e^{i\varphi}|1s\rangle\right), \qquad (2)$$

where $\varepsilon_{1s}$ and $\varepsilon_{2p}$ are the field-free eigenvalues, $\Delta E$ is the energy shift that depends on the strength of $F_{THz}$, and the hybridization is parameterized by an amplitude term ($\sin\beta$) and phase ($\varphi$) terms (see SM Section S4).

Due to the selection rule, $|2p, 0\rangle$ is not directly accessible by the absorption spectrum, but $|2p, -1\rangle$ can be measured under the THz-field dressing. Considering that the 1s-2p transition energy, $\hbar\Delta = \varepsilon_{2p} - \varepsilon_{1s}$, is ~63 meV (~15.2 THz) at room temperature [39], $|2p, -1\rangle$ is estimated to be ~5.8 THz below $|1s, 0\rangle$ when $F_{THz}$ is weak. Experimentally, the absorption feature of $|2p, -1\rangle$ can be resolved by the difference spectrum $\Delta\alpha(\omega, F_{THz}) \equiv \alpha(\omega, F_{THz}) - \alpha_0(\omega)$ [Fig. 1(c)] [14,40]. In Fig. 1(d), the energies $\varepsilon + \Delta E$ of $|1s, 0\rangle$ and $|2p, -1\rangle$ under different $F_{THz}$ are summarized, which can be well reproduced by the simulations (the solid lines) using Eqs. (1-2), indicating resonant contributions of the dressed excitonic states in the spectral range of 430 - 460 THz.

*Quantum-path interferometry.* – Spectro-temporal quantum-path interferometry is employed to extract the amplitude and phase information of strong-field-dressed excitons [Fig. 2(a)]. A weak NIR probe $F_{pr}(\omega, \tau)$, with $\tau$ representing the time delay between the THz and probe pulses, induces polarization corresponding to the $n^{\text{th}}$ order sideband, denoted by $P_{\omega,n}(\omega + n\Omega, \tau) = \chi(\omega + n\Omega)F_{pr}(\omega,\tau)$. Here, $\chi(\omega+n\Omega)$ is the effective susceptibility given by

$$\chi(\omega + n\Omega) = \sum_{l \in \mathbb{Z}}^{\varepsilon_m \in \text{FBZ}} \frac{V_{0m}^{l-n}(V_{0m}^{l})^*}{\varepsilon_m - \varepsilon_0 - \hbar(l\Omega+\omega) - i\Gamma_{\text{ex}}^n} \,, \qquad (3)$$

where $V_{0m}^{l}$ denotes the Fourier component of the transition dipole $\langle m, l|r|0\rangle$ with $m \in \{1s, 2p\}$, and $\Gamma_{\text{ex}}^n$ is a phenomenological $n$-dependent dephasing rate of the dressed excitons (see SM Section S4).

As illustrated in Fig. 2(a), when the probe bandwidth $\Delta\omega$ is larger than $2\Omega$, two quantum paths with equal energy lead to the transitions from $|0\rangle$ to two degenerate states: $|\omega_1 + n\Omega\rangle$ and $|\omega_2 + (n + 2)\Omega\rangle$, where $\omega_1$ and $\omega_2$ fall within the bandwidth of $\Delta\omega$ and are separated by $2\Omega$. The interference between these paths results in the oscillation of $I_{HSE}$ as a function of $\tau$ with a period of $\pi/\Omega$. The intensity oscillation can be described by

$$I_{HSE}(\bar{E}, \tau) = \left|P_{\omega_1,n}^R(\bar{E}, \tau)\right|^2 + \left|P_{\omega_2,n+2}^{NR}(\bar{E}, \tau)\right|^2$$
$$+ 2\left|P_{\omega_1,n}^R(\bar{E}, \tau)\right|\left|P_{\omega_2,n+2}^{NR}(\bar{E}, \tau)\right|\cos[2\Omega\tau + \Delta\phi(\bar{E}) + \Delta\varphi_{n,n+2}(\bar{E})], \qquad (4)$$

where $\bar{E} = \omega_1 + n\Omega = \omega_2 + (n + 2)\Omega$, $P_{\omega_1,n}^R$ and $P_{\omega_2,n+2}^{NR}$ respectively denote the polarization corresponding to the resonant (*R*) and non-resonant (*NR*) transitions respectively excited by the $\omega_1+n\Omega$ and $\omega_2+(n+2)\Omega$ quantum paths, $\Delta\phi = \phi(\omega_1) -$ $\phi(\omega_2)$ is the phase difference induced by $F_{pr}$, and $\Delta\varphi_{n,n+2} = \varphi_n - \varphi_{n+2}$ denotes the dipole phase difference between the two quantum paths. The first two terms on the right-hand side represent the non-oscillating contributions, while the third term leads to the $2\Omega$ oscillation of $I_{HSE}$. Detailed derivations are provided in SM Section S4.

The QPI scenario in Fig. 2(a) becomes evident when examining the $I_{HSE}$ oscillations as a function of $\Delta\omega$. The experiments were conducted on a 50-nm thick WSe$_2$ sample to avoid exciton resonance (see below). In Fig. 2(b), with $\Delta\omega \approx 95$ THz (> $4\Omega$), fully overlapped spectra of the neighboring HSE orders exhibit $2\Omega$ oscillation of $I_{HSE}$ across the entire spectral range, consistent with previous work [33]. However, at reduced $\Delta\omega$ (~65 THz, with $2\Omega<\Delta\omega<4\Omega$, spanning 335 to 400 THz), both the oscillating and non-oscillating spectral regions are simultaneously observed [Fig. 2(c)]. Remarkably, the spectral regions exhibiting $2\Omega$ oscillation precisely align with the regions where the quantum paths overlap (see SM Section S6).

Our results highlight the quantum nature of transient HSE dynamics [41,42]. This quantum mechanical interpretation is important, because it offers a unique opportunity to access the phase of coherently driven states. In particular, when one of the interfering quantum paths resonantly couples a dressed quantum state [as illustrated by $P^R_{\omega_1,n}$ in Fig. 2(a)], the resulting interferogram can provide important information about the time-domain dipole response. While the similar idea has been successfully implemented in laser-driven gas atoms [43,44] and solids [45,46], these studies have mainly focused on the high-energy region. Applying this concept to the low-energy intra-excitonic dynamics remains unexplored.

*Dephasing of strong-field-dressed excitons.* – In Fig. 3, we plot the spectro-temporal interferogram measured on the MoS$_2$ sample. The experimental THz-field envelope is shown in Fig. 3(a). To cover the exciton resonance, the $F_{pr}$ spectrum spans from 330 to 465 THz. The observed 2Ω oscillation results from the overlapping $n=0$ and 2 quantum paths (see SM Section S7). Consequently, the $I_{HSE}$ oscillation carries the dipole phase difference of $\Delta\varphi_{0,2}$. To illustrate the phase shifts more clearly, we remove the non-oscillating contributions by appropriate frequency filtering (see SM Section S5). Fig. 3(b1) shows the phase shifts of 2Ω oscillation for $\bar{E}$ in the range between 430 and 460 THz when the time delay $\tau$ is between -260 and -220 fs, corresponding to weak dressing fields. This energy range aligns with the resonant energies of $|1s, 0\rangle$ and $|2p, -1\rangle$ at the low-field limit [Fig. 1(d)]. Notably, as $\tau$ approaches the pulse temporal center ($\tau$ between -40 and 0 fs), where $F_{THz}$ reaches the maximum, the phase shifts in the same spectral range gradually diminish [Fig. 3(b2)]. The complete interferogram showing this evolution is provided in SM Fig. S11. Correspondingly, Fig. 3c shows two peaks in the experimental oscillation amplitudes within this spectral range.

The experimental phase variations are further supported by a temporally resolved analysis [Fig. 3(d)]. Specifically, for each sampled delay time, we perform a Fourier transform in a temporal window of 25 fs and the spectral range of 435 - 460 THz to extract the 2Ω-component phase. When $\tau$=-270 fs, we observe a robust phase shift $\Delta\varphi_{0,2}$ of -1.8 rad at $\bar{E} \approx 445$ THz, returning to 0 at $\bar{E} \approx 460$ THz. In stark contrast, the experimental phase shift vanishes when $\tau$= -10 fs. Notably, the phase results at $\tau$=-10 fs demonstrate that the contribution of probe-pulse dispersion ($\Delta\phi$) is negligible here.

According to Eq. (3), the phase shifts and amplitude peaks can be attributed to the $F_{pr}$-induced resonant transitions from $|0\rangle$ to $|1s, 0\rangle$ and $|2p, -1\rangle$ [Fig. 3(e)]. Whenever $F_{pr}$ induces a resonant transition, a $\pi$ phase shift arises, and an "S"-shape phase shift emerges when consecutive resonant transitions occur in the spectral range. Correspondingly, the spectral amplitude peaks around the resonant energies.

The diminish of phase variations under the strong driving field is attributed to the increased dephasing rate ($\Gamma_{ex}$). When $\Gamma_{ex}$ becomes much larger than the energy separation between the two resonant states, the twisting phase shifts become no longer observable due to the dephasing-induced spectral smearing. From a different perspective, the observation of phase variations under weak driving fields here indicates that $\Gamma_{ex}$ must be less than the energy separation between $|1s, 0\rangle$ and $|2p, -1\rangle$ ($|\hbar\Omega - \hbar\Delta| \approx 10$ THz), corresponding to a dephasing time ($T_2$) greater than 100 fs.

In Fig. 4(a), we summarize the field-strength dependence of $\Gamma_{ex}$. The experimental $\Gamma_{ex}$ was obtained by fitting the delay-dependent phase variations using a model that considers multiple resonant transitions (see SM Section S4). The field strength $F_{THz}$ was extracted at different time delays of the THz transients with varying peak-field strength, $F_{peak}$. The results indicate that the dephasing time ($T_2$) is ~300 fs ($\Gamma_{ex} \approx 3.0$ THz), when the THz field is absent, consistent with previous static studies where exciton-phonon scattering is the main dephasing channel [47–49].

An intriguing observation is the significant increase in the dephasing rate when $F_{THz}$ exceeds a threshold strength ($F_{th}$) of ~2.1 MV/cm. Below this threshold, we find that the dephasing rate $\Gamma_{ex}$ is lower than 10 THz ($T_2 > 100$ fs). Notably, the threshold strength ($F_{th}$≈2.1 MV/cm) has already reached the non-perturbative region (see the HSE intensity measurement in SM Section S2) [28,29,32,33], indicating a long dephasing time under a strong driving field.

To understand our results, we explore three potential exciton dephasing mechanisms [Fig. 4(b)]: multi-particle scattering ($\Gamma_{mult}$) [32], exciton-phonon scattering ($\Gamma_{ph}$) [26], and field-induced dissociation ($\Gamma_{diss}$) [50,51]. The total dephasing rate is given by $\Gamma_{ex} = \Gamma_{mult} + \Gamma_{ph} + \Gamma_{diss}$. The multi-particle and phonon environments can be theoretically evaluated using non-Markovian semiconductor Bloch equations [52] (see SM Section S8). The dephasing rate due to exciton dissociation is given by $\Gamma_{diss} = \frac{\gamma_{diss}}{2}$, where $\gamma_{diss}$ is the dissociation rate. Assuming the hydrogen-atom-like behavior, $\gamma_{diss}$ is obtained from the imaginary part of the eigenvalues in the Kramers-Henneberger frame [53] (also see SM Section S8).

Among different dephasing mechanisms, $\Gamma_{ph}$ is time- and field-strength independent, representing a residual value under low THz-field conditions. We find only $\Gamma_{diss}+\Gamma_{ph}$ can capture the threshold behavior shown in Fig. 4(a). Our model indicates that the THz field induces an energy upshift in the stable 1s state (nonionizing), resulting in a dramatic increase in the dissociation rate when the THz photon energy ($\hbar\Omega$) aligns with the resonant transition to the continuum state [Fig. 4(c)]. In contrast, due to the accumulation effect of the THz-induced carrier density, the rise of $\Gamma_{mult}$ is much slower compared to $\Gamma_{diss}$. Therefore, our findings suggest a field-strength-dependent exciton dephasing rate, which is further supported by the reemergence of the phase twists at positive time delays when the field-strength is reduced (see SM Section

S7). In our simulations [Figs. 3(d, f and g)], the same field-strength-dependent dephasing rate of $\Gamma_{diss}+\Gamma_{ph}$ is employed, which exhibits excellent agreement with the experimental results.

Here, our exciton-dissociation model predicts a threshold-field strength of ~1.2 MV/cm, which is lower than the experimental value [Fig. 4(a)]. This discrepancy could be attributed to the experimental uncertainties in measuring THz field strength within the material (see SM Section S1 and S2). Furthermore, the dielectric screening of the material may also cause uncertainty in the field strength. Despite these uncertainties, our exciton-dissociation model can clearly reproduce the threshold behavior [Fig. 4(a)], which is the key to distinguish among various mechanisms.

*Conclusion.* – Our study employs QPI within the framework of Floquet theory to reveal the field-strength-dependent dephasing mechanism of excitons in bulk $MoS_2$ under intense THz fields. By analyzing the field-strength- and time-dependence of the dephasing rates, we identify that exciton dissociation is the primary cause of strong dephasing at high driving fields, and we observe a long dephasing time (>100 fs) of an exciton under a strong driving field ($F_{THz}\approx 2.1$ MV/cm). These findings demonstrate the feasibility of strong-field Floquet engineering in dissipative quantum materials.

**Acknowledgments**

We wish to thank Yuan Wan for valuable discussion. Z. T. acknowledges financial support from the National Key Research and Development Program of China (Grant Nos. 2021YFA1400200 and 2022YFA1404700), the National Natural Science Foundation of China (Grant Nos. 12221004, 12274091 and 11874121) and the Shanghai Municipal Science and Technology Basic Research Project (Grant No. 22JC1400200). R. L. acknowledges financial support from the National Key Research and Development Program of China (Grant No. 2022YFA1604301), the National Natural Science Foundation of China (Grant No. 11974185). S. J. acknowledges the financial support from National Natural Science Foundation of China (Grant No. 12304378). Y. Z. acknowledges financial support from National Key Research and Development Program of China (Grant Nos. 2022YFA1403301 and 2018YFA0305600), Strategic Priority Research Program of Chinese Academy of Sciences (Grant No. XDB30000000), and Shanghai Municipal Science and Technology Commission (Grant No. 2019SHZDZX01). Z. T. is also thankful for the support from the Alexander-von-Humboldt foundation.

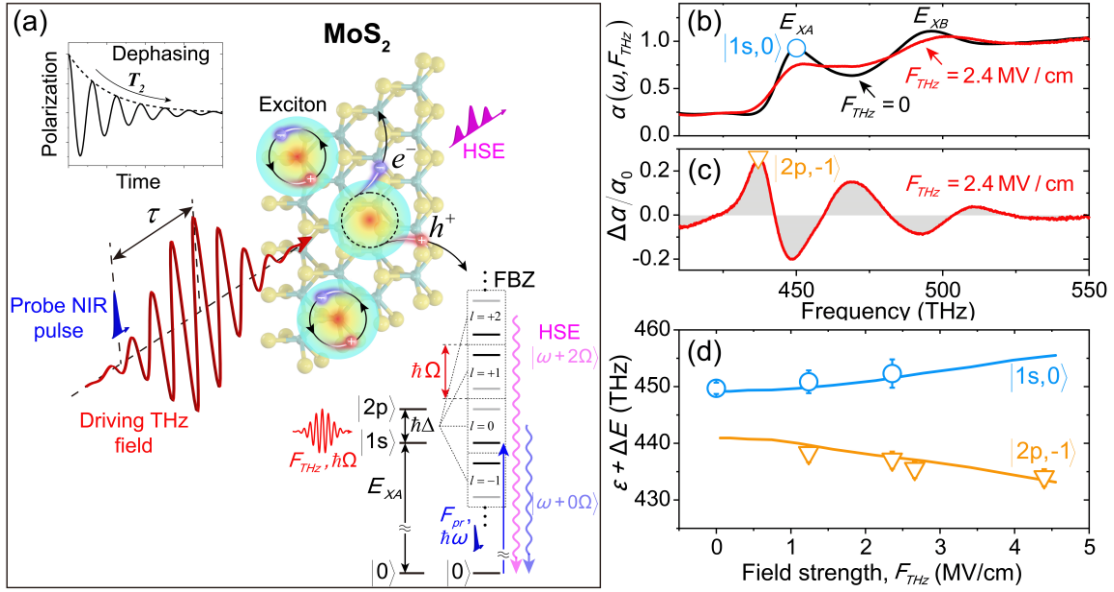

**Figure 1. (a)** Experimental setup: A strong THz field is applied to drive the excitons in bulk MoS$_2$ and the dephasing time ($T_2$) is probed by HSE with a NIR probe pulse arriving at a time delay of $\tau$. **<u>Inset:</u>** Illustration of the Floquet ladder of a THz-driven exciton, showing replicas of excitonic states with different dressed photon numbers $l$ in each FBZ. NIR pulse $F_{pr}$ probes the Floquet ladders through HSE. **(b)** Absorbance spectra of bulk MoS$_2$ with and without THz-field excitation. The absorption peak of $|1s, 0\rangle$ is labeled. **(c)** Normalized difference spectrum of absorbance for $F_{THz}$=2.4 MV/cm. The absorption peak of $|2p, -1\rangle$ is labeled. **(d)** Summary of the $|1s, 0\rangle$ and $|2p, -1\rangle$ energies as a function of $F_{THz}$.

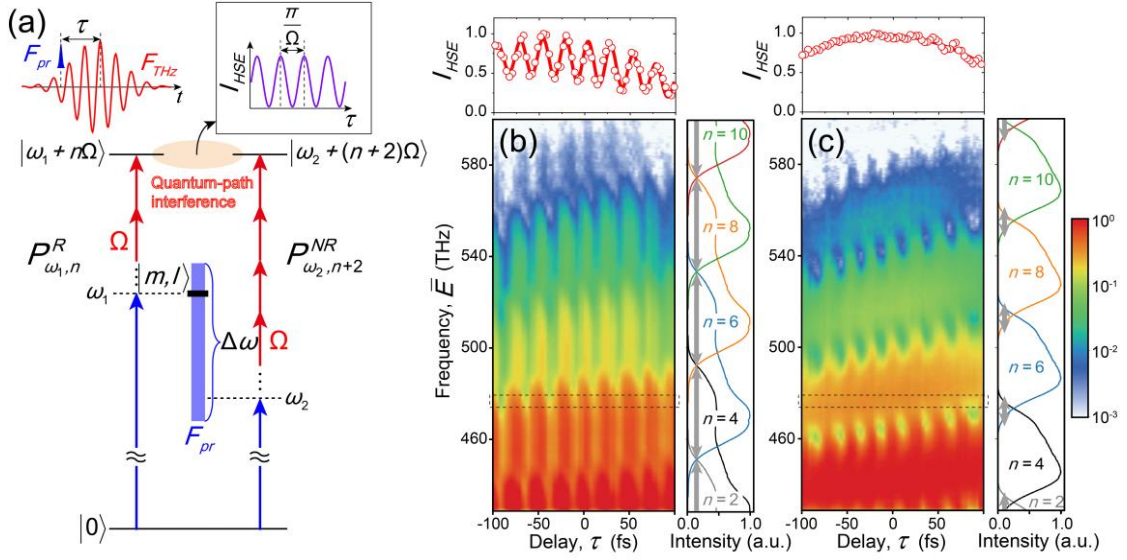

**Figure 2. (a)** Illustration of QPI between HSE photon states $|\omega_1 + n\Omega\rangle$ and $|\omega_2 + (n+2)\Omega\rangle$. The quantum path of $|\omega_1 + n\Omega\rangle$ is on resonance with a dressed excitonic state $|m, l\rangle$, carrying information about the corresponding transition dipole. QPI results in the $I_{HSE}$ oscillation as a function of $\tau$ with a period of $\pi/\Omega$. **(b)** Spectro-temporal interferogram obtained from a 50-nm WSe$_2$ sample with $F_{pr}$ bandwidth $\Delta\omega$ ~95 THz. **Right panel:** 1D lineouts of $F_{pr}$ spectra, upshifted by $n\Omega$ to illustrate the spectral regions for different HSE orders. Grey arrows show the quantum-path-overlapping regions. **Upper panel:** 1D lineout of $I_{HSE}$ in the selected spectral range. **(c)** Similar to **(b)**, but with narrower bandwidth of $\Delta\omega$~65 THz.

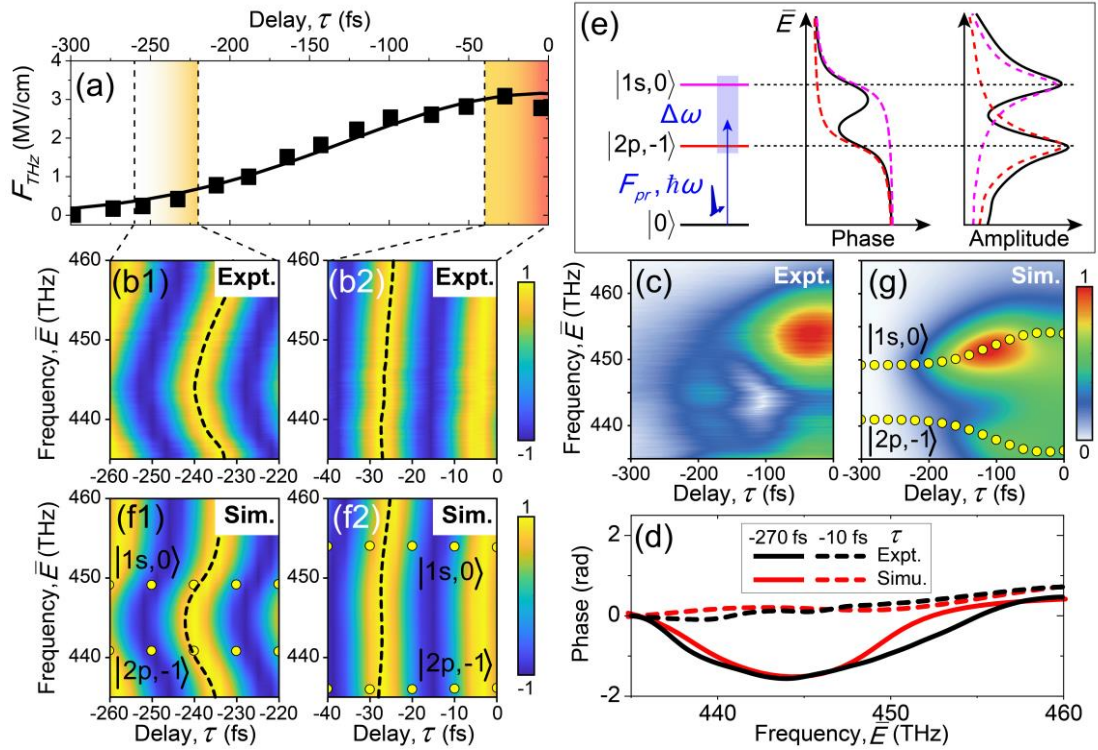

**Figure 3. (a)** Envelope of the THz field. **(b)** Experimental 2Ω components driven by a peak-field strength of $F_{peak} \approx 3$ MV/cm. **(b1)** The 2Ω component for $\tau$ between -260 and -220 fs. The dashed line highlights phase shifts in this spectral region. **(b2)** Similar to **(b1)**, for the 2Ω component between -40 and 0 fs. **(c)** Amplitudes of the 2Ω components extracted from the experimental results. **(d)** Spectral phase of 2Ω components at different $\tau$ extracted from the experimental and simulation results using the temporally resolved analysis. **(e)** Illustration of phase variations and oscillation amplitudes induced by resonant transitions to $|1s, 0\rangle$ and $|2p, -1\rangle$. Dashed lines represent individual resonant transitions, solid black lines depict combined results. **(f1-f2).** The 2Ω components extracted from numerical simulations, with yellow dots labeling shifts of $|1s, 0\rangle$ and $|2p, -1\rangle$ energies. **(g).** Amplitudes extracted from the numerical simulations.

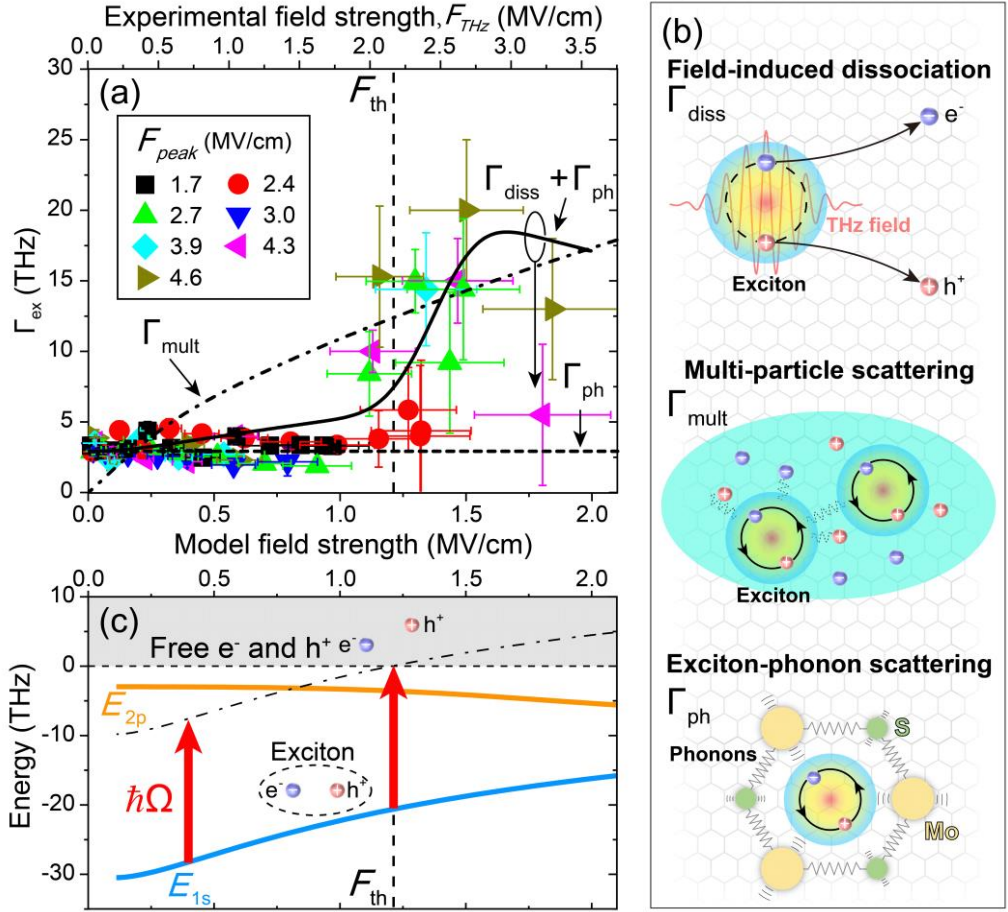

**Figure 4. (a)** Summary of dephasing rates, $\Gamma_{ex}$, under different THz-field strengths, $F_{THz}$. Predictions from different mechanisms are shown for comparison, with the solid line representing $\Gamma_{diss}+\Gamma_{ph}$, the dashed line for $\Gamma_{ph}$ and the dash-dot line for $\Gamma_{mult}$. Error bars for $\Gamma_{ex}$ result from fitting errors, while those for $F_{THz}$ stem from the experimental uncertainties. **(b)** Illustration of different dephasing mechanisms affecting THz-driven excitons. **(c)** Model results of the eigen-energies of 1s and 2p ($E_{1s}$ and $E_{2p}$) as a function of $F_{THz}$. Red arrows represent the transitions induced by a THz photon ($\hbar\Omega$). The dashed-dot line illustrates the upshift of $E_{1s}$ by $\hbar\Omega$.